\documentclass[twocolumn,preprintnumbers,amsmath,amssymb,prb]{revtex4}
\usepackage{graphicx}
\usepackage{dcolumn}
\usepackage{bm}

\begin{document}
\title{Impurity induced spin gap asymmetry in nanoscale graphene}
\author{Julia Berashevich and Tapash Chakraborty}
\email{chakrabt@cc.umanitoba.ca}
\affiliation{Department of Physics and Astronomy, University of 
Manitoba, Winnipeg, Canada, R3T 2N2}

\begin{abstract}
We propose a unique way to control both bandgap and the magnetic 
properties of nanoscale graphene, which might prove highly beneficial
for application in nanoelectronic and spintronic devices. We 
have shown that chemical doping by nitrogen along a single 
zigzag edge breaks the sublattice symmetry of graphene. This leads to 
the opening of a gap and a shift of the molecular orbitals localized on 
the doped edge in such a way that the spin gap asymmetry, which can 
lead to half-metallicity under certain conditions, is obtained.
The spin-selective behavior of graphene and tunable spin gaps help us 
to obtain semiconductor diode-like current-voltage characteristics, 
where the current flowing in one direction is preferred over the other. 
The doping in the middle of the graphene layer results in an impurity 
level between the HOMO and LUMO orbitals of pure graphene (again, much like in semiconductor 
systems) localized on the zigzag edges thus decreasing the bandgap and adding unpaired electrons, 
and this can also be used to control graphene conductivity.
\end{abstract}
\maketitle

\section{Introduction}
Applications of graphene with its unique physical properties \cite{nov1,expt2,tapash,tapash1,david} 
in nanoelectronics \cite{chen,kern}, 
magnetism and spintronics \cite{cohen,rudb,esq,cho,karpan}, hang 
crucially on its bandgap and spin ordering at the zigzag edges. 
A bandgap can be opened in graphene by breaking the certain 
symmetries. For example, interaction of graphene with its
substrate, such as SiC, leads to the charge exchange between 
them which breaks the sublattice symmetry \cite{zhou}. Moreover, 
the quantum confinement effect also has been found to introduce 
a small bandgap in graphene nanoribbons \cite{kim}, just as was 
predicted earlier theoretically \cite{cohen1,nak,lee,pisani}. 
The effect of bandgap opening and spin ordering between the zigzag edges 
are found to be directly linked to each other \cite{harrison}. 
When the spins align along the zigzag edges and spin states 
localized at opposite edges have the same spin orientation, 
then symmetry of graphene is preserved and the
system is gapless. Otherwise, if the spin-up states are localized 
along one zigzag edge and the spin-down along the other, the 
sublattice symmetry is broken which leads to a gap. 
In the light of a recent breakthrough 
in fabrication of nanoribbons of required size through unzipping 
of carbon nanotubes, the nanoribbons and nanoscale graphene are the most 
promising systems for application in nanoelectronics \cite{nature}.

Manipulation of the spin ordering is important for both 
graphene magnetism and its electronic properties. There are several 
approaches which have been proposed recently to  
control the spin ordering along the edges \cite{fer,my,gun,bouk,li,hod,kan}. 
One of them is the termination of the zigzag edges by functional 
groups \cite{fer}. This has the advantage that one can 
achieve half metallicity in this process. However, there are some 
serious issues involved here. Firstly, many of these functional 
groups are placed out of the graphene plane thus making the whole 
structure non-planar and, most importantly, termination was applied 
to every second edge cell, which makes its technological application 
very difficult. In fact, we found that the strong interactions of 
the graphene lattice with some of the functional groups, such as 
NH$_2$ and NO$_2$, lead to buckling of the graphene layer and 
twisting of the functional groups, which subsequently may result 
in the disappearance of the half-metallicity of graphene.
The curling of the graphene layer has been seen in 
earlier studies as well \cite{kan}, where the boundary conditions
were found to control the graphene planarity, namely 
the curling occurs for stand-alone systems.
For nanoscale graphene the ferromagnetic ordering
of the spin states along the zigzag edges can be also
achieved as subsequence of adsorption of gas and
water molecules on the graphene surface, as we have shown
in our previous study \cite{my}. The adsorption leads to pushing 
of the $\alpha$- and $\beta$-spin states to the opposite zigzag 
edges thereby breaking sublattice symmetry and opening a gap.
In some cases the spin asymmetry can occur.
For example, the adsorption of HF gas molecule provide
the HOMO-LUMO gap of 2.1 eV for $\alpha$-spin state and
of 1.2 eV for $\beta$-spin state. However, due to the weak
interaction between adsorbant and graphene surface the
phenomena of the spin alignment along the edges takes place
locally, thus limiting its application.

The connection of the phenomena of bandgap opening and of the spin 
ordering with the sublattice symmetry lead us to conclude that 
breaking of this symmetry is the main direction to achieve the 
required semiconductor-type bandgap in graphene and a tunable
spin ordering. Here we make a proposal that the symmetry 
breaking can be done by chemical doping along a single zigzag edge.
This method is far superior to the earlier approaches involving 
edge termination by functional groups because doping can be 
done for every unit cell along the zigzag edge and thus preserve the planarity of graphene. 
Doping not only breaks the graphene symmetry, but also can induce the spin gap asymmetry 
due to the energetic shift of the molecular orbitals localized on the doped edge.
In a structure with broken symmetry, the HOMO$_{\alpha}$ and LUMO$_{\beta}$ 
orbital states are localized at one edge, while HOMO$_{\beta}$ 
and LUMO$_{\alpha}$ are at the other. Suppose the doping shifts  
the HOMO$_{\alpha}$ and LUMO$_{\beta}$ orbital states
localized at one edge down, then the HOMO$_{\alpha}$-LUMO$_{\alpha}$ 
bandgap ($\Delta_{\alpha}$) is increased, while $\Delta_{\beta}$, 
in contrast, will be reduced.   
If a certain type of impurities can 
cause a significant shift of the bands, then the half-metallicity of 
graphene may occur. This is what we set out to investigate here. 
An important advantage of this approach is that we expect insensitivity 
of spin selective behavior to the quality of the edges, when the band 
shift induced by the impurities is stronger than the contribution from 
the edge defects. We also investigate the possibility of obtaining an 
impurity level in the middle of the graphene bandgap by doping (in
analogy to semiconductors), which has a lot of technological 
implications as well. Our study of nanoscale graphene was based on 
the quantum chemistry methods using the spin-polarized density 
functional theory with the semilocal gradient corrected functional 
(UB3LYP/6-31G) performed in the Jaguar 7.5 program \cite{jaguar}.  

\section{Symmetry of nanoscale graphene}

\begin{figure}
\includegraphics[scale=0.45]{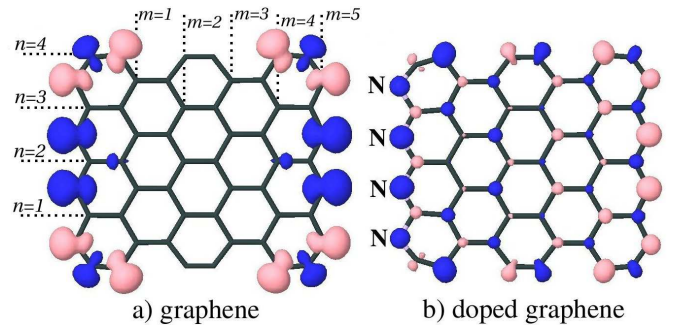}
\caption{\label{fig:fig1} (color online). The spin density 
distribution: (a) for nanoscale graphene optimized with the D2h 
point-group symmetry, and (b) for the case when one edge is 
doped by nitrogen, where the highest symmetry is the planar 
symmetry. Different colors indicate the $\alpha$-- (light) 
and $\beta$--spin (dark) states. The spin density is plotted 
for isovalues of $\pm0.01$ e/\AA$^3$. The $n$ and $m$ are introduced to identify the  
the number of the carbon rings along the zigzag edge ($n$) and 
along the armchair edge ($m$).}
\end{figure}

Bulk graphene has hexagonal symmetry, while the highest possible 
symmetry of nanoscale graphene would be the D2h planar 
symmetry with an inversion center. The D2h 
symmetry results in structurally identical corners exhibiting 
ferromagnetic ordering of the spin-polarized states localized at the corners,
as presented in Fig.~\ref{fig:fig1} (a). 
According to the NBO (natural bond orbital)
analysis, the localized electrons at the corners are 
unpaired $sp$ electrons belonging to
non-bonded orbitals, which are located 
at the bottom of conduction band or top of the valence band.
For this symmetry, both $\alpha-$ and $\beta$-spin
states of the HOMO and LUMO orbitals are localized on the zigzag edges but 
their spin density is equally distributed between two edges. 
For nanoscale graphene of D2h symmetry the HOMO-LUMO gap
appears due to confinement and edge effects \cite{cohen1}.
The degeneracy of the $\alpha$- and $\beta$-spin states belonging 
to the HOMO and LUMO orbitals depends on the edge 
configuration, i.e., $\alpha$- and $\beta$ states can be 
non-degenerate or degenerate depending on number of the carbon 
rings along the zigzag and armchair edges \cite{my}.
The degeneracy reappears for large structures, such as $n\ge 8$, $m\ge 7$ 
(see notation in Fig.~\ref{fig:fig1} (a)).
The increase of the graphene size leads to 
disapperance of the confinement effect and as a result closing of the gap.
Thus, for $n=4$ and $m=5$ the gap is $\sim$0.5 eV and already 
for $n=6$ and $m=7$ the gap is suppressed to $\sim$0.19 eV.
The influence of the confinement effect on the 
graphene gap has been already confirmed experimentally \cite{kim}.

However, the state of D2h symmetry is not the ground state for 
nanoscale graphene. Graphene, optimized with C2v symmetry, where the
mirror plane of symmetry is parallel to the armchair edges, 
has a total energy lower than that for the D2h symmetry. 
For the C2v symmetry, the HOMO and LUMO orbitals 
are characterized by the $\alpha-$ and $\beta$-spin
states localized on the opposite zigzag edges.
Because the carbon atoms at the opposite zigzag edges belong to 
different sublattices, such spin distribution breaks sublattice 
symmetry and opens a gap ($\sim 1.63$ eV for $n=4$ and $m=5$).
The size of the gap is comparable to that 
found for nanoribbons \cite{harrison}. 
The large gap of nanoscale graphene obtained here 
is a result of significant contribution of the confinement effect, 
as the nanoscale graphene is confined in all directions.

\begin{figure*}
\includegraphics[scale=0.80]{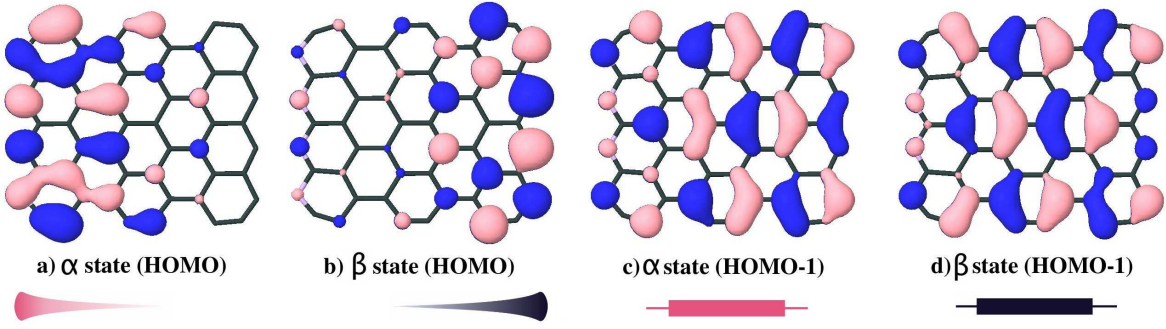}
\protect\caption{\label{fig:fig2} (color online). Spin polarizations in nanoscale 
graphene where the left edge is doped by nitrogen. 
Different colors correspond to different signs of 
the molecular orbital lobes. The electron densities are plotted 
for isovalues of $\pm0.02$ e/\AA$^{3}$:
(a) $\alpha$-state of HOMO ($E_{\rm HOMO}=-6.04$ eV) 
(b) $\beta$-state of HOMO ($E_{\rm HOMO}=-5.43$ eV)
(c) $\alpha$-state of (HOMO-1)($E_{\rm HOMO-1}=-6.38$ eV) 
(d) $\beta$-state of (HOMO-1) ($E_{\rm HOMO-1}=-6.43$ eV).
The HOMO and LUMO are found to be localized at 
the single zigzag edges (edge states), while (HOMO-1) and (LUMO+1) -- 
delocalized over the entire graphene structure (surface states). Bottom pictures 
show the representation of the localized and surface states.}
\end{figure*}

The localization of the $\alpha-$ and $\beta$-spin states belonging
to HOMO and LUMO at the opposite zigzag edges
is important for the application of graphene in spintronics \cite{cohen,rudb,hod1,dutta}.
However, the C2v symmetry state is a highly metastable state.
Its total energy is comparable to that of D2h symmetry with a difference of $\sim -0.5$ eV 
for small structures such as $n=4$ and $m=5$, 
but the difference decreases exponentially down to $\sim -0.02$ eV 
with increasing the structure size up to $n\ge 6$ and $m\ge 7$, 
that has good agreement with earlier work \cite{lee}, and disappears
when $n>8$ and $m>8$ .
The competition of the C2v state
with C1 symmetry, which is not constrained to have 
spin ordering along the zigzag edge, 
is even more crucial because of almost identical magnitude of their total energy.
However, we found that the distortion or dissimilarity induced 
along a single zigzag edge not only breaks the sublattice symmetry of the graphene, but
can control the spin ordering, thereby stabilizing the ground state of the C2v symmetry. 
The highest possible symmetry of the doped graphene is lowered from 
D2h to C2v symmetry as a result of the edge dissimilarity. The spin density 
distribution for the nanoscale graphene with one zigzag edge doped by 
nitrogen is presented in Fig.~\ref{fig:fig1} (b). 
The localized states along the zigzag edges are formed by 
unpaired electrons belonging to the
natural non-bonded orbitals, which participate in formation of HOMO and LUMO orbitals.
The $\alpha$-- and $\beta$--spin states of HOMO and LUMO orbitals
are spatially separated, i.e. localized at opposite zigzag edges. The (HOMO-1) 
and (LUMO+1) orbitals usually correspond to the surface states, 
redistributed over the entire graphene structure. The surface states 
are important for conductivity of graphene in a transverse electric 
field, because the charge transfer between the spatially separated 
HOMO and LUMO orbitals may occurs through participation of the 
surface states. The electron density distribution for the edge states 
and the surface states is presented in Fig.~\ref{fig:fig2}. The slight 
difference between the $\alpha$- and $\beta$-spin surface states 
(HOMO-1) is due to doping of the left zigzag edge.
The $\alpha$- and $\beta$-spin 
states remain spatially separated with increasing structure 
size. 

\section{Half-metallicity of graphene}
The edge dissimilarity allows us to explore the required 
properties, such as the semiconductor-type bandgap and 
localization of $\alpha$-- and $\beta$--spin states at 
opposite zigzag edges. Moreover, the spatial separation 
of the $\alpha$-- and the $\beta$--spin states resulted 
from doping of the single zigzag edge is stable in comparison 
to the water adsorption \cite{my}.
Doping of a single edge shifts the band energies of the 
orbitals which are strongly localized at this edge. Such 
a shift provides an opportunity to obtain another useful 
property which is important for spintronics -- the 
half-metallicity of graphene. For the HOMO or LUMO orbitals, 
which are shown to be localized at the edges, doping can 
create a strong non-degeneracy of the $\alpha$-- and 
$\beta$--spin states, because these states are spatially 
separated and localized at the opposite edges. Moreover, 
the HOMO$_{\alpha}$ and LUMO$_{\beta}$ orbitals are 
localized at one edge, while HOMO$_{\beta}$ and LUMO$_{\alpha}$ 
at the other. If doping increases 
the bandgap $\Delta_{\alpha}$ for the $\alpha$-spin 
state, then the bandgap $\Delta_{\beta}$ for the 
$\beta$-spin state, in contrast will be reduced, and vice 
versa. Therefore, doping induces the spin gap asymmetry in graphene. 
Materials exhibiting asymmetric gaps for the $\alpha$- and $\beta$-spin 
states where one gap is of semiconductor type while the other is an
insulator, are known as half-semiconductor materials, but  
if one of them is metallic, the system is half-metallic. 
Therefore, by choosing the right doping we can 
achieve a stable half-metallicity in graphene which will 
be an important step forward for applications in spintronics.

\begin{figure*}
\includegraphics[scale=0.85]{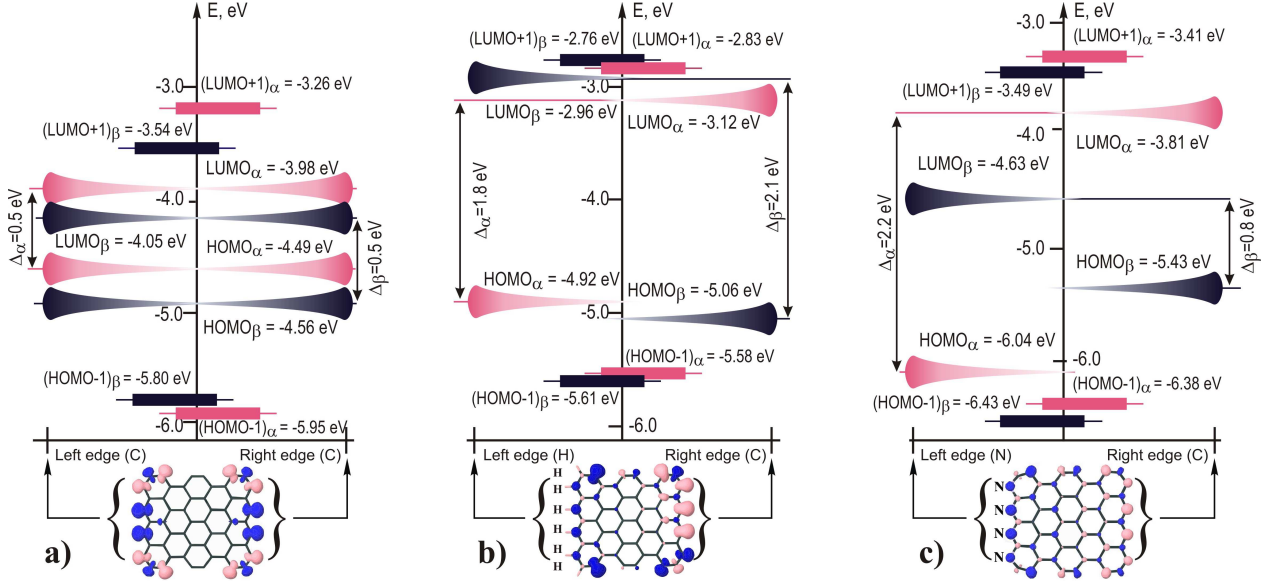}
\protect\caption{\label{fig:fig3}(color online).
Schematic diagrams showing the distribution of the 
edge states and surface states in the energy scale and over the graphene structure 
(see the bottom pictures in Fig.~\ref{fig:fig2} for pictorial description of the states). 
The structures at the bottom demonstrate 
the spin distribution with isovalues of $\pm0.01$ e/\AA$^{3}$.
(a) nanoscale graphene optimized with the D2h symmetry,
(b) with left edge terminated by hydrogen and (c) 
with the left zigzag edge doped by nitrogen. For the localized states the energy 
levels (HOMO and LUMO) show density distribution (schematic), 
particularly delocalization of the orbitals between the two 
edges if the D2h symmetry is preserved, and their localization 
on the zigzag edges when sublattice symmetry is broken and 
C2v symmetry becomes to be highest possible symmetry. 
The surface states ((HOMO-1) and (LUMO+1)) 
are delocalized over the entire graphene structure (see for 
example Fig.~\ref{fig:fig2} (c,d)).}
\end{figure*}

We have investigated the transformation of the 
electronic structure of nanoscale graphene 
due to the induced edge dissimilarities. The results 
are schematically presented in Fig.~\ref{fig:fig3}. For 
nanoscale graphene with the D2h symmetry, a small bandgap 
occurs due to the confinement effect. The HOMO and LUMO 
orbitals in this case are localized at the zigzag edges, but 
their electron density is equally redistributed over both 
edges (see Fig.~\ref{fig:fig3} (a)). 
Termination of the left 
zigzag edge by hydrogen (see Fig.~\ref{fig:fig3} (b)) opens a gap as a result 
of breaking of the sublattice symmetry, thereby lowering D2h symmetry to a stable 
ground state of C2v symmetry. 
The hydrogenation 
leads to saturation of the dangling $\sigma$ bonds
at the terminated edge but does not significantly change the 
energy of the HOMO$_{\alpha}$ and LUMO$_{\beta}$ states 
localized at this edge. The resulting non-degeneracy 
of the $\alpha$- and $\beta$-spin states is not large, and 
the HOMO-LUMO gap of the $\alpha$-spin state ($\Delta_{\alpha}$
=1.8 eV) is almost identical to that of the $\beta$-spin state 
($\Delta_{\beta}$=2.1 eV). The doping of the left zigzag edge 
by nitrogen (see Fig.~\ref{fig:fig3} (c)) shifts down the 
orbital energies of the HOMO$_{\alpha}$ and LUMO$_{\beta}$ 
states localized at the doped edge and results in a strong 
non-degeneracy of the orbitals. This leads to a slight 
enhancement of the HOMO-LUMO gap for the $\alpha$-spin state 
up to $\Delta_{\alpha}$=2.2 eV, but significantly decreases 
the HOMO-LUMO gap for the $\beta$-spin state down to $\Delta_{\beta}$
=0.8 eV. The length of the nitrogen-carbon bond at the edges 
is found to be $d_{N-C}$=1.35 \AA, which is similar to the 
carbon-carbon bonds $d_{C-C}$=1.39 \AA. Similar results are 
obtained for phosphorus impurities, where the gaps are 
$\Delta_{\alpha}$=2.0 eV and $\Delta_{\beta}$=0.9 eV.
Phosphorus is, however, less useful because of the large phosphorus-carbon 
bond ($d_{P-C}$=1.78 \AA) which can lead to destruction of 
the lattice.
We have also investigated the possibility to 
dope the single zigzag edge of the nanoscale graphene by 
other impurities, such as oxygen and boron, but they are not as
effective as nitrogen. The oxygen doping leads to strong delocalization 
of the electron density of the orbitals localized at the edges.
The doping by boron leads to even more troubles due to the long 
boron-carbon bonds at the edges ($d_{B-C}$=1.42 \AA) and shifting of the 
states localized at the edges from the HOMO-LUMO gap 
deeper into the conduction and valence bands.

For the nanoscale graphene structure investigated in this work,
the spin asymmetry is achieved but bandgap magnitude for 
$\alpha$- and $\beta$-spin states corresponds to the 
half-semiconductor behavior ($\Delta_{\alpha}$=2.2 eV,$\Delta_{\beta}$
=0.8 eV). Increasing the size of the graphene  
results in a decrease of both the $\Delta_{\alpha}$ and $\Delta_{\beta}$ 
gaps due to the diminishing of the confinement effect. 
Therefore, for graphene structures doped by nitrogen
or phosphorus of size $n\ge 6$ and $m\ge 7$, 
the $\Delta_{\beta}$ gap is closer to metallic type.
Thus, for $n= 6$ and $m= 7$ the gap for $\alpha$-spin
state is suppressed down to 1.13 eV while for $\beta$-spin state
down to 0.19 eV, which corresponds to the {\it half-metallic behavior 
of graphene}. An external electric 
field applied between the zigzag edges has been shown 
\cite{cohen,rudb,hod1,dutta} to shift the band of graphene with 
spatially separated and degenerated $\alpha$- and $\beta$-spin states. The 
electric field shifts the bands in such a way that for the 
$\alpha$-spin the HOMO and LUMO levels move closer to each other 
in the energy scale, while for $\beta$-spin they move apart. At a 
certain electric field $\Delta_{\alpha}$ vanishes, thereby 
creating a metallic behavior of graphene. 
If $+ E_{c}$ electric field can close the bandgap for 
the $\alpha$-spin state, the $- E_{c}$ leads to bandgap disappearance 
for the $\beta$-spin state. Therefore, the current 
voltage characteristic of such a structure will be symmetrical 
because the $\Delta_{\alpha}$ equals $\Delta_{\beta}$ and for both spin states the 
switch from the semiconductor behavior to metallic occurs at the 
same critical electric field $\pm E_{c}$. 
The advantage of graphene with spin gap asymmetry, i.e. different 
$\Delta_{\alpha}$ and $\Delta_{\beta}$ gaps, found in this work 
is the different values of the critical electric field required 
to close these gaps, such that $\mid E_{c(\beta)}\mid < \mid 
E_{c(\alpha)}\mid$ when $\Delta_{\beta} < \Delta_{\alpha}$. 
Therefore, this structure will be characterized by the spin-polarized 
current and by a non-symmetric current-voltage characteristics 
as for a semiconductor diode, when the current flow in one 
direction is preferable to the other.

\section{Doping of graphene}

\begin{figure}
\includegraphics[scale=0.30]{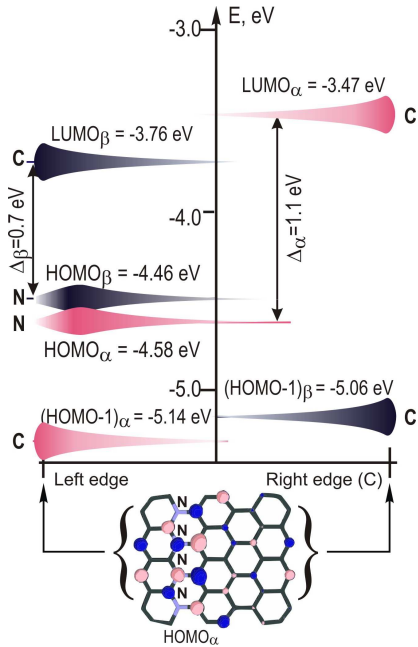}
\protect\caption{\label{fig:fig4}(color online).
Schematic diagram showing the distribution of the 
edge states (LUMO, HOMO-1) and states localized by the dopant 
in the middle of the graphene structure (HOMO) 
(see the bottom pictures in Fig.~\ref{fig:fig2} for pictorial description of the states). 
The HOMO$_{\alpha}$ and HOMO$_{\beta}$ are extra impurity levels that appear due 
to the doping and replaces the occupied orbital localized on the left carbon edge by 
shifting it dipper into the valence band. The inset picture 
(in brackets) demonstrates the electron density distribution for 
the HOMO$_{\alpha}$ with isovalues of $\pm0.01$ e/\AA$^{3}$.}
\end{figure}

We have also investigated the influence of impurities on the 
electronic structure of graphene in the case when they are not
embedded at the zigzag edges. Replacing carbon atoms by 
nitrogen atoms in a graphene lattice results in the appearance 
of impurity levels {\it inside} of both the $\Delta_{\alpha}$ 
and $\Delta_{\beta}$ gaps. The energy diagram of localization 
of the molecular orbitals for the doped graphene is presented 
in Fig.~\ref{fig:fig4}. As we mentioned earlier, in pure graphene 
the HOMO and LUMO orbitals are localized at the zigzag edges. 
The applied doping creates one extra occupied orbital (HOMO) 
which is localized at the embedded nitrogen atoms and located 
above the occupied orbital belonging to edges, which becomes 
HOMO-1. The NBO analysis has shown that this extra orbital 
is formed by the unpaired $sp$ electron localized on each 
nitrogen atoms. This reduces the 
HOMO-LUMO gap ($\Delta_{\alpha}$=1.1 eV and $\Delta_{\beta}$=0.7 
eV) while preserving the spin asymmetry. 

In an applied in-plane electric 
field the charge transfer occurs between the orbitals localized 
on the opposite zigzag edges, i.e., in our case such
transfer occurs between the HOMO-1 and LUMO orbitals, which is a multi-step process
with participation of HOMO.
Because the gap is decreased and each nitrogen atom adds an unpaired electron 
into the system due to the doping, the conductivity of 
graphene would be significantly enhanced, and can be controlled 
as it is done in semiconductor devices.

\section{Conclusion}

We have investigated the possibility to control the 
electronic and magnetic properties of nanoscale graphene. 
We found that if pure graphene can be characterized by a small bandgap 
and no spin ordering at the zigzag edges the dissimilarity of the 
edges induced by doping impurities lowers the highest possible symmetry 
to C2v, which is characterized by the spin ordering along the zigzag edges and 
their antiparallel alignment between opposite zigzag edges.
Moreover, impurities embedded at a single zigzag edge shifts in the energy scale the 
molecular orbitals localized at this edge, thereby decreasing the 
bandgap for one spin channel and increasing the other. Under these 
conditions, the half-metallic behavior can be achieved. 
Nitrogen doping in the middle of the graphene surface is found 
to have the prospect for application in nanoelectronics due to the 
appearance of the occupied impurity levels in the bandgap. 
The impurity level results in a decrease of the bandgap 
of $\sim$ 2.0 eV by one half and contains unpaired electrons, which should lead to an 
enhancement of the conductivity. Therefore, both the conductivity 
of the nanoscale graphene and its magnetic properties can be 
controlled by the impurities.

\section{Acknowledgments}
The work was supported by the Canada Research Chairs
Program and the NSERC Discovery Grant.\\

\end{document}